# Probing millicharged particles with ultrasensitive optical nonlinear sensor based on levitated cavity optomechanics

Jian Liu and Ka-Di Zhu*

*Key Laboratory of Artificial Structures and Quantum Control (Ministry of Education),*

*School of Physics and Astronomy, Shanghai Jiao Tong University,*

*800 DongChuan Road, Shanghai 200240, China,*

*Collaborative Innovation Center of Advanced Microstructures, Nanjing, China*

**Abstract:** Particles with electric charge $10^{-12}$e in bulk mass are not excluded by present experiments. In the present letter we provide a feasible scheme to measure the millicharged particles via the optical cavity coupled to a levitated microsphere. The results show that the optical probe spectrum of the micro-oscillator presents a distinct shift due to the existence of millicharged particles. Owing to the very narrow linewidth($10^{-7}$Hz) of the optical Kerr effect, this shift will be more obvious, which makes the millicharges more easy to be detectable. We propose a method to eliminate the polarization force background via the homogeneously charged ring, which makes the scheme displays strong advantages in precision than the current experiments. The technique proposed here paves the way for new applications for probing dark matter and nonzero charged neutrino in the condensed matter.

## Contents

## I. INTRODUCTION

Millicharged particles(MCPs) or epsilon charged particles are hypothetical particles with an electric charge $e^{'} = \epsilon e$, which is much smaller than the elementary charge e, namely $\epsilon \ll 1$. The existence of the particles with small, unquantized electric charge can be introduced into the standard model in a variety of ways[1]. There has been interest in the possibility of a small, nonzero electric charge for the neutrino[2,3], and the possibility that dark matter carries a fractional or epsilon charge has been considered recently[4-6]. The search for the epsilon charges for the range from keV to GeV progressed through the efforts of many direct experiments and indirect observations over many years, but the evidence is still lacking[7-11]. Now the search for stable particles with millicharges in bulk matter is still on the initial stage. Previous research using magnetic levitation electrometers or Millikan oil drop techniques in matter did not have sensitivity less than 0.1e[12-16]. Recently, Moore et.al report the search for particles with $\epsilon \geq 10^{-5}$ using optically levitated microspheres[17]. Based the optical trapping scheme, we propose a levitated sensor to detect the epsilon charge in micro-scale with improved sensitivity of $\epsilon \sim 10^{-12}$ in this paper by the cavity pump-probe technology.

Cavity optomechanics, which explores the interaction between optical field and mechanical motion, has witnessed rapid advances in recent years[18]. One of the principal advantages of cavity-optomechanical systems is the built-in readout of mechanical motion via the light field transmitted through (or reflected from) the cavity. Various of schemes have been proposed for the measuring of the parametric displacement[19], mechanical Fock-state detection[20], optical feedback cooling[21] and extremely sensitive force sensing[22,23]. Recently, the optical pump-probe technique has become a popular topic, which offers an effective method to study the light-matter interaction. The optical pump-probe technology includes a strong pump field and a weak probe field. The strong pump laser is used to stimulate the system to generate coherent optical effect, while the weak laser plays the role of probe beam. Therefore, the linear and nonlinear optical effects can be observed on the probe spectrum in the two field's control scheme.

*zhukadi@sjtu.edu.cn

Most recently, this pump-probe scheme has also been realized experimentally in optical cavity[24-26]. In the present paper, we apply the pump-probe fields to the cavity-microsphere coupling system and study theoretically the optical nonlinear in the optomechanical system. The results show that an ultra narrow linewidth of the enhanced peak can be achieved through the high vacuum allowed. Based on the frequency shift of the Kerr peak, the detection of signature of MCPs in bulk matter will be carried out by the transmission spectrum. In view of it, we expect that the optomechanical detector could improve the search for epsilon charges via an enhanced sensitivity of $10^6 \sim 10^7$ than current experiments[17]. At last we discuss the limits of the thermo-mechanical noise on the sensitivity in our scenario.

## II. MODELS

As shown in Fig.1, a silica microshpere with radius $a = 0.5\mu m$ is placed inside a Fabry-Perot cavity and optically trapped at the antinodes of the optical lattice. Then we radiate a strong pump field(with frequency $\omega_p$) and a weak probe(with frequency $\omega_{pr}$) field on the microsphere-cavity system. The dielectric object is subjected to an electric field generated by a homogeneously charged ring with radius $R = 50\mu m$.

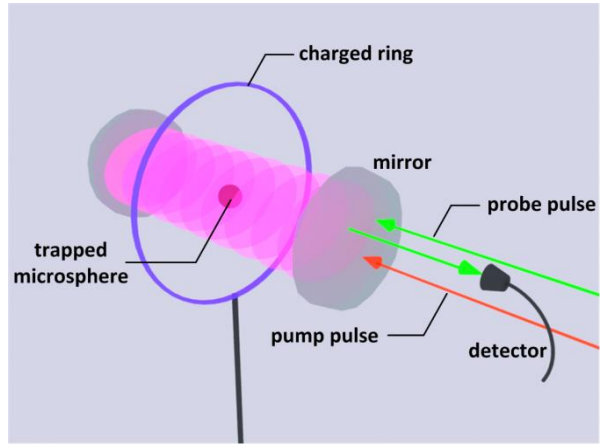

Fig.1 Schematic diagram of the proposed setup for detecting millicharged particles by optical pump-probe technology in the cavity. Our approach involves optically trapping a fabricated microsphere in the antinode of an optical standing wave. To eliminate the polarization force background, we use a homogeneously charged ring to produce symmetrical electric field near the microsphere and the microsphere should be trapped at the center of the ring.

Let us denote the cavity annihilation (creation) operator by $\sigma(\sigma^+)$ with the commutation relation $[\sigma, \sigma^+] = 1$. The bosonic operators of the microsphere oscillator are represented by $s$ and $s^+$ with $[s,s^+] = 1$. Then the Hamiltonian of the whole system in a rotating frame reads as follow[27-29]:

$$H = \hbar\Delta_c\sigma^+\sigma + \hbar\omega_n s^+s - \hbar g\sigma^+\sigma(s^+ + s)$$
$$-i\hbar\Omega_p(\sigma - \sigma^+) - i\hbar\Omega_{pr}(\sigma e^{i\Delta_{pr}t} - \sigma^+ e^{-i\Delta_{pr}t}), \quad (1)$$

where g is the optomechanical coupling rate, $\omega_n$ is the vibrational frequency of the microsphere, $\omega_c$ is the frequency of the cavity mode, $\Delta_c = \omega_c - \omega_p$ is pump-cavity detuning. We also introduce the Rabi frequency of the pump field and the probe field inside the cavity $\Omega_p = \sqrt{2P_p\kappa/\hbar\omega_p}$ and $\Omega_p = \sqrt{2P_{pr}\kappa/\hbar\omega_{pr}}$, where $P_p$ is the pump power, $P_{pr}$ is the power of the probe field, respectively. We then define the operator $n = s^+ + s$ and the detuning of the probe and pump field $\Delta_{pr} = \omega_{pr} - \omega_p$. The temporal evolutions of $\sigma$ and n can be obtained and the quantum Langevin equations are given by adding the damping terms. In the present case, we distinguish between the channels associated with the input coupling (decay rate $\kappa_{ex}$) and the other loss processes (overall decay rate $\kappa_0$, including loss through the second mirror).

$$\frac{d\sigma}{dt} = -(i\Delta_c + \kappa)\sigma + ign\sigma + \Omega_p + \Omega_{pr}e^{-i\Delta_{pr}t} + \sqrt{\kappa_{ex}}\hat{a}_{in} + \sqrt{\kappa_0}\hat{u}_{in}, \quad (2)$$

$$\frac{d^2n}{dt^2} + \gamma_n\frac{dn}{dt} + \omega_n^2 n = 2\omega_n g\sigma^+\sigma + \hat{\xi}(t), \quad (3)$$

where $\kappa = \kappa_0 + \kappa_{ex}$ is the total energy damping rate of the cavity, $\gamma_n$ is the damping rates of the vibrational mode of the microsphere. The input field $\hat{a}_{in}$ should be thought of as a stochastic quantum field. The same kind of description holds for the "unwanted" channel associated with $\hat{u}_{in}$. The noise correlators associated with the input fluctuations are given by

$$\langle \hat{a}_{in}(t)\hat{a}_{in}^+(t')\rangle = \delta(t-t'), \quad (4)$$
$$\langle \hat{a}_{in}(t)\hat{a}_{in}^+(t')\rangle = 0. \quad (5)$$
$$\langle \hat{u}_{in}(t)\hat{u}_{in}^+(t')\rangle = \delta(t-t'), \quad (6)$$
$$\langle \hat{u}_{in}(t)\hat{u}_{in}^+(t')\rangle = 0. \quad (7)$$

The motion of microsphere are affected by thermal bath of Brownian and non-Markovian stochastic process[30]. The quantum effects on the resonator are only displayed in the limit of very high quality factor. The resonator mode of levitated particles are affected by a Brownian stochastic force with zero mean value, and $\hat{\xi}(t)$ has the correlation function [31]

$$\langle \hat{\xi}^+(t)\hat{\xi}(t')\rangle = \int \frac{\kappa\omega d\omega}{2\pi\omega_n} e^{-i\omega(t-t')}\left[1 + \coth\left(\frac{\hbar\omega}{2k_B T}\right)\right]. \quad (8)$$

Here $k_B$ is Boltzmann's constant and T is the effective resonator temperature. Following standard methods from quantum optics, we derive the steady-state solutions to Eqs.(2) and (3) by setting all the time derivatives to zero,

$$\bar{\sigma} = \frac{\Omega_p}{(i\Delta_c + \kappa) - ign}, \qquad \bar{n} = \frac{2g|\sigma|^2}{\omega_n}. \quad (9)$$

We can always rewrite each Heisenberg operator as the sum of its steady-state mean value and a small fluctuation with zero mean value as follows,

$$\sigma = \sigma_0 + \delta\sigma, \quad n = n_0 + \delta n. \quad (10)$$

Inserting these equations into the Langevin equations, one can safely neglect the nonlinear terms such as $\delta\sigma\delta n$. Since the driving fields are weak, we will identify all operators with their expectation values, and drop the quantum and thermal noise terms. Then the linearized Langevin equations can be written as

$$\langle\delta\dot{\sigma}\rangle = -\kappa\langle\delta\sigma\rangle + ig(n_0\langle\delta\sigma\rangle + \sigma_0\langle\delta n\rangle) + \Omega_p + \Omega_{pr}e^{-i\Delta_{pr}t}, \quad (11)$$

$$\langle\delta\ddot{n}\rangle + \gamma_n\langle\delta\dot{n}\rangle + \omega_n^2\langle\delta n\rangle = 2\omega_n g\sigma_0^2, \tag{12}$$

To solve these equations, we make the ansatz as follows:

$$\langle\delta\sigma\rangle = \sigma_+ e^{-i\Delta_{pr}t} + \sigma_- e^{i\Delta_{pr}t}, \tag{13}$$

$$\langle\delta n\rangle = n_+ e^{-i\Delta_{pr}t} + n_- e^{i\Delta_{pr}t}. \tag{14}$$

Upon substituting these equations into Eqs.(11) and (12), and upon working to the lowest order in $\Omega_p$ but to all orders in $\Omega_{pr}$, we can obtain in the steady state

$$\sigma_- = \frac{G}{(P+\kappa^2)}\frac{\Omega_{pr}\Omega_p^2}{B(Y^2-P^2)+2PG\omega_0}. \tag{15}$$

Here $\sigma_-$ is a parameter in analogy with the Kerr nonlinear optical susceptibility with $Y = -i\Delta_c + \kappa$, $P = i\Delta_c - ign_0$, $B = -\Delta_{pr}^2 - i\gamma_n\Delta_{pr} + \omega_n^2$, $G = 2ig^2\omega_n$, $n_0 = 2g|\sigma_0|^2/\omega_n$ and $\omega_0$ can be resolved by

$$\Omega_p^2 = \omega_0\left[\kappa^2 + \left(\Delta_c - \frac{g^2\omega_0}{\omega_n}\right)^2\right]. \tag{16}$$

Using the input-output relation $\sigma_{out}(t) = \sigma_{in}(t) - \sqrt{2\kappa}\sigma(t)$, the amplitude of the Kerr nonlinear can be expressed as $\sigma_{out-} = -\sqrt{2\kappa}\sigma_-$. The transmission intensity of the output field is then given by[24]

$$|T_{out-}|^2 = 2\kappa\left|\frac{\sigma_{out-}}{\Omega_{pr}}\right|^2. \tag{17}$$

### III. MEASURMENT

For numerical work, we use a silicon microsphere with the density of $\rho = 2.3 \times 10^3 \text{kg/m}^3$ and the dielectric constant $\varepsilon = 2$. The trapping frequency $\omega_n$ depends on the trapping beam intensity, it can be modulated by the intracavity power[27]:

$$\omega_n = \frac{1}{2\pi}\left(\frac{6K^2 I_0}{\rho c}\text{Re}\frac{\varepsilon-1}{\varepsilon+2}\right)^2, \tag{18}$$

here c is the speed of light, $K = 2\pi/\lambda$ is the wave numbles of the trapping beams with wavelength $\lambda = 1\mu m$. In what follows, we choose the intensity $I_0 = 10^{-2} W/\mu m^2$, hence one can obtain $\omega_n = 150 kHz$.

The quality factor one would get for a levitated microsphere is solely determined by the air molecule impacts. Random collisions with residual air molecules provide the damping $\gamma_n = \omega_n Q^{-1}$ and thus, the quality factor due to the gas dissipation can be defined as[27,32]

$$Q = \frac{m\omega_n v}{pA} = \frac{a\rho\omega_n v}{3p}. \tag{19}$$

Here $v = \sqrt{k_B T/m_{gas}}$ is the thermal velocity of the gas molecules, p the gas pressure and A the surface area of the resonators. For the microsphere under the pressure $p = 10^{-10}$ Torr, we get $Q = 1.2 \times 10^{12}$ in the room temperature(T=300K). This value is considered to be within reach of moderate experimental improvements[27,33].

The optomechanical coupling between the cavity field and the microsphere can be described by the coupling rate[27]

$$g = \frac{3V_s}{4V_c}\frac{\varepsilon-1}{\varepsilon+2}\omega_c, \tag{20}$$

where $V_s$ and $V_c$ are the microsphere and the optical mode volumes, respectively. We use the cavity parameters from a recent experiment on the observation of optically levitation[34],

$\omega_c \sim 10^{15}$ Hz, the length L=1cm and mode waist w = 75μm ($V_c = (\pi/4)Lw^2$). Then the optomechanical coupling rate g can be obtained as g$\sim 10^6$ Hz.

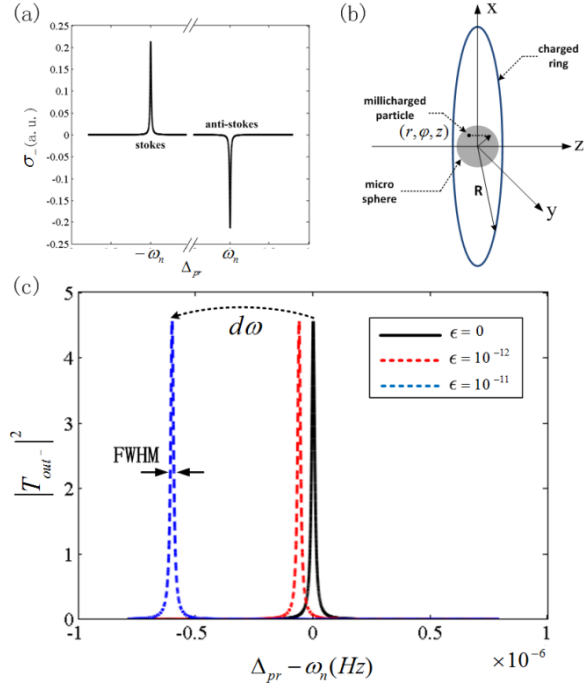

Fig.2 (a)The plot of absorption spectrum as a function of probe-pump detuning. The unit of the Stokes scattering absorption intensity is an arbitrary unit(a.u.) which is a relative unit of measurement to show the ratio of amount intensity to a predetermined reference measurement. (b)Force analysis for the millicharged particles in the microsphere. (c)Nonlinear optical spectrum of the probe field as a function of probe-pump detuning before (black solid line) and after (red dashed line and blue dash line) the action between millicharge and electric field gradient. The signals of the MCP with $\epsilon = 10^{-12}$ and $\epsilon = 10^{-11}$ can be well recognized in the spectrum. Other parameters used are $\Omega_p = 1$kHz, $\Delta_c = 0$, $Q = 10^{12}$.

In order to remove the net charge of levitated sensor in trapping, a fiber-coupled xenon flash lamp is used to illuminate the electrode surfaces near the microsphere[17]. To determine the original frequency, we consider the case that there is no applied electric field around the microsphere. We take the pump-cavity detuning $\Delta_c = 0$. According to Eqs.(15)(16), we depict the Kerr nonlinear susceptibility $\sigma_-$ as a function of probe-pump detuning $\Delta_{pr}$ in Fig.2(a). We can find that two sharp Kerr peaks are located at the original frequencies of the microsphere resonator on the spectrum. The physical origin can be understood as four-wave mixing (FWM), an important third-order nonlinear process. Such a phenomenon is attributed to the destructive quantum interference effect between the microsphere mode and the beat of the two optical fields via the cavity. If the beat frequency of two lasers $\Delta_{pr}$ is close to the resonance frequency of microsphere, the mechanical mode starts to oscillate coherently, which results in Stokes-like ($\Delta_{pr} = \omega_n$) and anti-Stokes-like ($\Delta_{pr} = -\omega_n$) scattering of light from the strong intracavity field.

For small oscillation amplitudes, the levitated particle experiences a harmonic trapping potential with a trap stiffness(or the spring constant) $k = \omega_n^2 m$, The effective spring constant can be defined as $k' = k + \nabla F$, here $\nabla F$ represents the force gradient at the equilibrium position of the levitated particle. If we consider the interaction between the electric field gradient and the single MCP, we get $\nabla F = \epsilon e \nabla E$, here $\nabla E$ is the electric field gradient. The resonance frequency $\omega_n$ can be converted into $\omega_n' = \sqrt{k + \nabla F}/\sqrt{m}$. If the differentiation is sufficiently small, we get

$\omega_n \cong \omega_n'$, and $\nabla F/2k \cong \omega_n/\omega_n' - 1$. Assume $d\omega = \omega_n' - \omega_n$, based on the frequency shift of the resonance mode, one can obtain

$$\nabla F = \frac{2k}{\omega_n} d\omega. \tag{21}$$

Here the frequency shift is dependent linearly on the force gradient via the linearized responsivity $2k/\omega_n$, which mean that the shift is the true signature of the force at the equilibrium position of microsphere.

To probe the tiny MCPs-induced frequency shift on the spectrum, then we electricize the ring homogeneously with a assumed linear density $\upsilon = 10^{-4} C/m$. The main systematic effect limiting the measurement sensitivity in Ref.[17] is the interacting primarily between the electric field and the dipole moments of the microspheres. In order to minimize this effect, we suggest using the homogeneously charged ring. The microsphere can be trapped at the centre of the ring, thereby the applied electric field is distributed symmetrically along the horizontal axis. Thus the resultant polarization force in the microsphere is zero. Now we consider a MCP appears in the microsphere with the position $(r, \phi, z)$ in cylindrical coordinates as shown in Fig.2(b). The microsphere will experience a force due to the electromagnetic interaction between the charged ring and the MCP. The electric field along z axis at the MCP's location can be obtained by

$$E_z = \frac{q}{4\pi^2 \varepsilon_0} \frac{2z}{\sqrt{R^2 + r^2 + z^2 + 2Rr}(R^2 + r^2 + z^2 - 2Rr)} O, \tag{22}$$

where R is the radius of the ring, $q = 2\pi R \upsilon$ is the total quantity of electric charge of the ring, $\varepsilon_0$ is the vacuum permittivity, $O = (\pi/2)[1 - \left(\frac{1}{2}\right)^2 U^2 - \left(\frac{1 \cdot 3}{2 \cdot 4}\right)^2 \frac{U^4}{3} - \cdots]$ is a total elliptical integral and the parameter $U^2 = 4Rr/(R^2 + r^2 + z^2 + 2Rr)$. Since the scale of the microsphere is much smaller than the ring, namely $r, z \ll R$, the electric field can be obtained within approximation,

$$E_z \approx \frac{q}{4\pi^2 \varepsilon_0} \frac{2z}{R^3}. \tag{23}$$

Then we can obtain the force gradient

$$\nabla F_z = \frac{2\epsilon e q}{4\pi^2 \varepsilon_0 R^3}. \tag{24}$$

We replace Eq.(21) in the equation for $\nabla F_z$ and considering $k = \omega_n^2 m$, it is easy to show that the single MPC will induce a shift as

$$d\omega = \frac{\epsilon e \upsilon}{2\pi \varepsilon_0 R^2 \omega_n m}. \tag{25}$$

Interestingly, we find that the frequency shift is independent with the position of the MPC in the microsphere in this case. That is to say, no matter where it appears, the shift is merely decided by the total electron charge $\epsilon e$.

Fig.2(c) shows the transmission spectrum of the probe laser $|T_{out-}|^2$ as a function of the probe detuning $\Delta_{pr} - \omega_n$ with ($\epsilon \neq 0$) and without ($\epsilon = 0$) the MPC. The black curve shows the result when there is no charges in the microshpere. As the MPC with $\epsilon = 10^{-12}$ appears in the matter, the transmission peak will present an distinct shift of $d\omega = 6 \times 10^{-7} Hz$ (see the red dash curve). The blue sharp peak shows that the enhanced peak will exhibit a larger frequency shift on the transmission spectrum as we consider the MPC of $\epsilon = 10^{-11}$ in the matter. The distance of shift $d\omega$ in the probe spectrum as a function of $\epsilon$ follows a linear relationship as $d\omega = 6 \times 10^5 \epsilon$. This may provide a direct method to measure the millicharges in this coupled system. The ability to improve the resolution depends on the full width at half maximum(FHWM) of the transmission peak. Considering the peak FHWM in Fig.2(c) is approximately $6 \times 10^{-7}$ Hz, we can calculate from Eq.(25) that the millicharges can be measured down to a precision of $\epsilon_{min} \sim 10^{-12}$ in this coupling system.

## IV. CONSTRAINTS AND LIMITS

The ultimate sensitive limits for nanomechanical resonators operating in vacuo are imposed by a number of fundamental physical noise processes. In our scheme, the thermal noise of the mechanical motion is the dominant noise source. In order to get the fundamental limits, we need to consider the minimum measurable frequency shift $\delta\omega_{min}$ that can be resolved in a realistic noise system. For the case of $Q \gg 1$[35],

$$d\omega_{min} \approx \left[\frac{k_B T \omega_n \Delta f}{E_c Q}\right]^{\frac{1}{2}}. \tag{26}$$

Here, $E_c = m\omega_n^2 \langle x_{rms}\rangle^2$ represents the maximum drive energy. According to the equipartition theorem, the root-mean-square(rms) amplitude of a trapped microsphere at thermal equilibrium is $\langle x_{rms}\rangle = \sqrt{k_B T/(m\omega_n^2)}$. The smallest measurement bandwidth $\Delta f$ is given by $\Delta f = \frac{2}{\gamma_n} = \frac{2\pi\omega_n}{Q}$. Considering the quality factor $Q = 10^{12}$, we can achieve that $d\omega_{min} = 4.9 \times 10^{-7}$ Hz at room temperature(T=300K). Since $d\omega_{min} \approx$ FHWM, we would expect that the claimed sensitivity for the millicharge can be resolvable in actual experiment in vacuum.

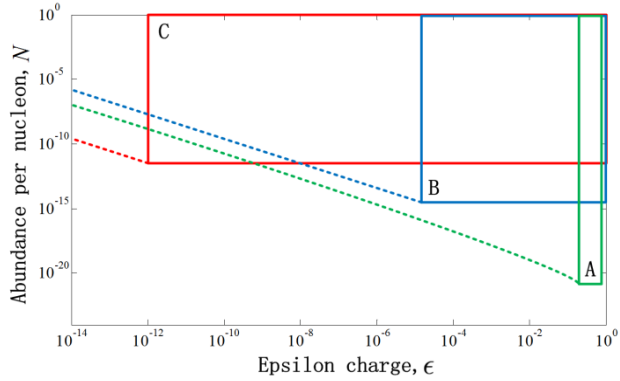

Fig.3 Constraints on the abundance of millicharged particles per nucleon N versus the epsilon charge. The results from this work (red rectangle C) are compared to previous results from levitated microspheres (blue rectangle B) [17] and magnetic levitometer (green rectangle A) [14] experiments. The lines extending from each region show the upper limits below the single particle threshold.

At last we conservatively calculate limits on the abundance per nucleon N of charged particles with $\epsilon$ in Fig.3, and compare to previous limits from optically levitated microspheres[17] and magnetic levitometer [14] experiments. The rectangle region A and B denotes the space directly excluded by the previous experiments above their single particle thresholds. The region C denotes the constraints for our scheme. As shown in Fig.3, the method we have described provides the direct search for MCP in condensed matter with sensitivity of $\epsilon \geq 10^{-12}$ to single particle. Over this full range, the upper limit on the abundance per nucleon is at most $N \sim 10^{-12}$. The method we have described promises to be the most sensitive approach for detecting millicharged particles. The lines extending from each region show limits on the abundance below the single particle threshold for comparison.

## V. CONCLUSION

In this work, we investigate the Kerr nonlinear spectrum in the optomechanical system where a microsphere coupled to a cavity. We have proposed an optical method to detect the existence of MCPs in bulk matter via pump-probe technology. A direct scheme to determine the electric charge is also demonstrated. The result shows that a sharp peak can be obtained at the resonant frequency. Based on the frequency shift of the vibrational mode, the MCPs can be measured on the spectrum. The Kerr nonlinearity with an ultra narrow linewidth gives rise to an improved sensitivity of order $10^{-12}e$, which is much smaller than that in current experiments and many other theoretical expectations. In the resonant millicharges detection, the strain sensitivity of our setup is limited by the thermal motion of the sensor. We would expect that the claimed sensitivity can be still achieved in actual noise system in the ultrahigh vacuum.

## ACKNOWLEDGMENTS

National Natural Science Foundation of China (11274230, 11574206); the Basic Research Program of the Committee of Science and Technology of Shanghai (14JC1491700).

**References**

1 : S. Davidson, S. Hannestad, and G. Raffelt, *Updated bounds on milli-charged particles*, J. High Energy Phys. **5**, (2000) 003.

2 : A. Y. Ignatiev and G. C. Joshi, *Neutrino electric charge and the possible anisotropy of the solar neutrino flux*, Phys. Rev. D **51**, (1995) 2411.

3 : R. Foot, G.C. Joshi, H. Lew and R.R. Volkas, *Charge quantization in the standard model and some of its extensions*, Mod. Phys. Lett. A **5**, (1990) 2721.

4 : J. M. Cline, Z. Liu, and W. Xue, *Millicharged atomic dark matter*, Phys. Rev. D **85**, (2012) 101302.

5 : E. Gabrielli, L. Marzola, M. Raidal, and H. Veermae, *Dark matter and spin-1 milli-charged particles*, J. High Energy Phys. **8**, (2015) 150.

6 : S. Profumo, *GeV dark matter searches with the NEWS detector*, Phys. Rev. D **93**, (2016) 055036.

7 : T. Mitsui, R. Fujimoto, Y. Ishisaki, Y. Ueda, Y. Yamazaki, S. Asai, and S. Orito, *Search for invisible decay of orthopositronium*, Phys. Rev. Lett. **70**, (1993) 2265.

8 : A. A. Prinz, R. Baggs et al, *Search for Millicharged Particles at SLAC,* Phys. Rev. Lett **81**, (1998) 1175.

9 : S. N. Gninenko, N. V. Krasnikov, and A. Rubbia, *New limit on millicharged particles from reactor neutrino experiments and the PVLAS anomaly*, Phys. Rev. D **75**, (2007) 075014.

10 : A. D. Dolgov, S. L. Dubovsky, G. I. Rubtsov, and I. I. Tkachev, *Constraints on millicharged particles from Planck data*, Phys. Rev. D **88**, (2013) 117701.

11 : A. Haasa, C. S. Hillb, E. Izaguirrec, and I. Yavin, *Looking for milli-charged particles with a new experiment at the LHC*, Phys. Lett. B **746**, (2015) 117.

12 : P. C. Kim, E. R. Lee, I. T. Lee, M. L. Perl, V. Halyo, and D. Loomba, *Search for Fractional-Charge Particles in Meteoritic Material*, Phys. Rev. Lett. **99**, (2007) 161804.

13 : I. T. Lee, S. Fan, V. Halyo, E. R. Lee, P. C. Kim, M. L. Perl, H. Rogers, D. Loomba, K. S.

Lackner, and G. Shaw, *Large bulk matter search for fractional charge particles*, Phys. Rev. D **66**, (2002) 012002.

14 : M. Marinelli and G. Morpurgo, *Searches of fractionally charged particles in matter with the magnetic levitation technique*, Phys. Rep. **85**, (1982) 161.

15 : P. F. Smith, *Searches for fractional electric charge in terrestrial materials*, Annu. Rev. Nucl. Part. Sci. **39**, (1989) 73.

16 : G. S. LaRue, J. D. Phillips, and W. M. Fairbank, *Observation of Fractional Charge of (1/3)e on Matter*, Phys. Rev. Lett. **46**, (1981) 967.

17 : D. C. Moore, A. D. Rider, and G. Gratta, *Search for Millicharged Particles Using Optically Levitated Microspheres*, Phys. Rev. Lett. **113**, (2014) 251801.

18 : M. Aspelmeyer, T. J. Kippenberg, and F. Marquardt, *Cavity optomechanics*, Rev. Mod. Phys. **86**, (2014) 1391.

19 : A. A. Clerk, M. H. Devoret, S. M. Girvin, F. Marquardt, and R. J. Schoelkopf, *Introduction to quantum noise, measurement, and amplification*, Rev. Mod. Phys. **82**, (2010) 1155.

20 : Z. Yin, T. Li, X. Zhang, and L. M. Duan, *Large quantum superpositions of a levitated nanodiamond through spin-optomechanical coupling*, Phys. Rev. A **88**, (2013) 033614.

21 : P. F. Barker and M. N. Shneider, *Cavity cooling of an optically trapped nanoparticle*, Phys. Rev. A **81**, (2010) 023826.

22 : A. A. Geraci, S. B. Papp, and J. Kitching, *Short-Range Force Detection Using Optically Cooled Levitated Microspheres*, Phys. Rev. Lett. **105**, (2010) 101101.

23 : J. Liu and K. D. Zhu, *Nanogravity gradiometer based on a sharp optical nonlinearity in a levitated particle optomechanical system*, Phys. Rev. D **95**, (2017) 044014.

24 : S. Weis, R. Rivière, S. Deléglise, E. Gavartin, O. Arcizet, A. Schliesser, and T. J. Kippenberg, *Optomechanically induced transparency*, Science **330**, (2010) 1520.

25 : J. D. Teufel, D. Li, M. S. Allman, K. Cicak, A. J. Sirois, J. D. Whittaker and R. W. Simmonds, *Sideband cooling micromechanical motion to the quantum ground state*, Nature **471**, (2011) 204.

26 : A. H. Safavi-Naeini, T. P. Mayer Alegre, J. Chan, M. Eichenfield, M. Winger, Q. Lin, J. T. Hill, D. E. Chang and O. Painter, *Electromagnetically induced transparency and slow light with optomechanics*, Nature **472**, (2011) 69.

27 : D. E. Chang, C. A. Regal, S. B. Papp, D. J. Wilson, J. Ye, O. Painter, H. J. Kimble, and P. Zoller, *Cavity opto-mechanics using an optically levitated nanosphere*, Proc. Natl. Acad. Sci. U.S.A. **107**, (2010) 1005.

28 : A. C. Pflanzer, O. Romero-Isart, and J. I. Cirac, *Master-equation approach to optomechanics with arbitrary dielectrics*, Phys. Rev. A **86**, (2012) 013802.

29 : C. Genes, D. Vitali, P. Tombesi, S. Gigan, and M. Aspelmeyer, *Ground-state cooling of a micromechanical oscillator: Comparing cold damping and cavity-assisted cooling schemes*, Phys. Rev. A **77**, (2008) 033804.

30 : C. Gardiner and P. Zoller, *Quantum Noise, 2nd ed.* (Springer, Berlin, 2000), p. 425.

31 : V. Giovannetti, and D. Vitali, *Phase-noise measurement in a cavity with a movable mirror undergoing quantum Brownian motion*, Phys. Rev. A. **63**, (2001) 023812.

32 : F. R. Blom, S. Bouwstra, M. Elwenspoek, J. H. J. Fiuitman, *Dependence of the quality factor of micromachined silicon beam resonators on pressure and geometry*, J. Vac. Sci. Technol. B **10**, (1992) 19.

33 : J. Gieseler, L. Novotny, and R. Quidant, *Thermal nonlinearities in a nanomechanical oscillator*, Nature Phys. **9**, (2013) 806.

34 : N. Kiesel, F. Blaser, U. Delic, D. Grass, R. Kaltenbaek, and M. Aspelmeyer, *Cavity cooling of an optically levitated submicron particle*, Proc. Natl. Acad. Sci. U.S.A. **110**, (2013) 14180.

35 : K. L. Ekinci, Y. T. Yang, and M. L. Roukes, *Ultimate limits to inertial mass sensing based upon nanoelectromechanical systems*, J. Appl. Phys. **95**, (2004) 2682.

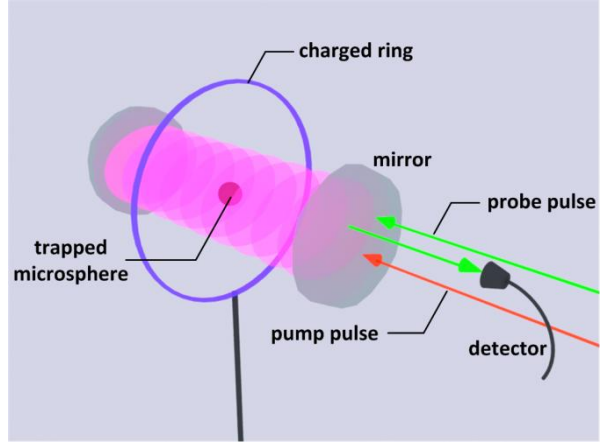

Fig.1 Schematic diagram of the proposed setup for detecting millicharged particles by optical pump-probe technology in the cavity. Our approach involves optically trapping a fabricated microsphere in the antinode of an optical standing wave. To eliminate the polarization force background, we use a homogeneously charged ring to produce symmetrical electric field near the microsphere and the microsphere should be trapped at the center of the ring.

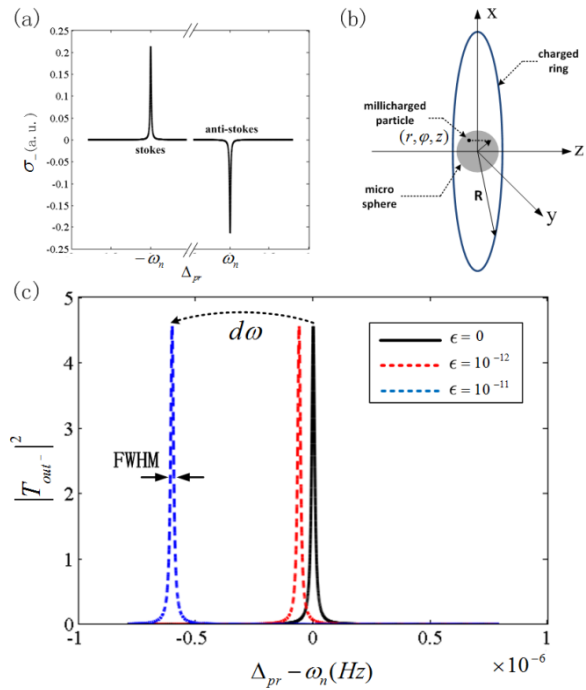

Fig.2 (a)The plot of absorption spectrum as a function of probe-pump detuning. The unit of the Stokes scattering absorption intensity is an arbitrary unit(a.u.) which is a relative unit of measurement to show the ratio of amount intensity to a predetermined reference measurement. (b)Force analysis for the millicharged particles in the microsphere. (c)Nonlinear optical spectrum of the probe field as a function of probe-pump detuning before (black solid line) and after (red dashed line and blue dash line) the action between millicharge and electric field gradient. The signals of the MCP with $\epsilon = 10^{-12}$ and $\epsilon = 10^{-11}$ can be well recognized in the spectrum. Other parameters used are $\Omega_p = 1\text{kHz}$, $\Delta_c = 0$, $Q = 10^{12}$.

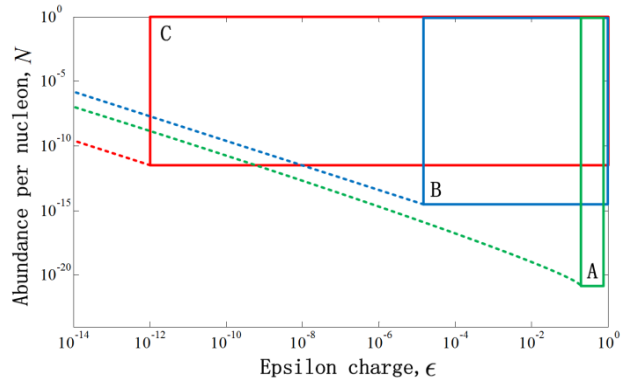

Fig.3 Constraints on the abundance of millicharged particles per nucleon N versus the epsilon charge. The results from this work (red rectangle C) are compared to previous results from levitated microspheres (blue rectangle B) [17] and magnetic levitometer (green rectangle A) [14] experiments. The lines extending from each region show the upper limits below the single particle threshold.